\documentclass[conference]{IEEEtran}
\usepackage{graphicx}
\usepackage{amsmath}
\usepackage{textcomp}
\usepackage{lineno,hyperref}

\ifCLASSINFOpdf
\else
\fi
\begin{document}
%

\title{{\large \textcopyright 2016 IEEE. Personal use of this material is permitted. Permission from IEEE must be obtained for all other uses, in any current or future media, including reprinting/republishing this material for advertising or promotional purposes, creating new collective works, for resale or redistribution to servers or lists, or reuse of any copyrighted component of this work in other works. DOI: 10.1109/MWSCAS.2016.7870083\\ .\\ I. Damaj, A Unified Analysis Approach for Hardware and Software Implementations, The $59^{th}$ IEEE International Midwest Symposium on Circuits and Systems, Abu Dhabi, UAE, 16\textendash19 October, 2016. P 577\textendash580.\\  \url{https://doi.org/10.1109/MWSCAS.2016.7870083}}\\ A Unified Analysis Approach for Hardware and Software Implementations}

\author{\IEEEauthorblockN{Issam W. Damaj}
\IEEEauthorblockA{Department of Electrical and Computer Engineering\\
American University of Kuwait\\
Salmiya, Kuwait\\
Email: idamaj@auk.edu.kw}}


%


\maketitle

\begin{abstract}
Modern computing systems are hybrid in nature and employ various processing technologies that range from specific- to general-purpose processors. In co-design environments, specific-purpose processors, also known as hardware, work to support software implementations under general-purpose systems to create high-performance computers. Algorithms and computationally intensive tasks are partitioned among the different processing subsystems to achieve desirable degrees of parallel processing and performance characteristics. In this paper, a unified statistical performance analysis formulation is presented. The proposed statistical formulation combines the heterogeneous characteristics of both hardware and software implementations to provide grounds for thorough evaluations. The formulation includes the development of performance profiles, key indicators, and the composition of a master indicator based-on heterogeneous measurements. The investigation includes a case-study that targets a set of simple cryptographic algorithms. The two main targeted high performance computing devices are multi-core processors for software implementations and high-end Field Programmable Gate Arrays for hardware implementations.  
\end{abstract}


%
\IEEEpeerreviewmaketitle

\section{Introduction}
Modern high-performance computers (\textit{HPCs}) are hybrids of multi-core processors, graphical processing units (\textit{GPUs}), high-density programmable logic devices (\textit{HDPLDs}), to name a few. Within hybrid systems, algorithms can be partitioned and distributed or fully-delegated to one subsystem. Hybrid \textit{HPCs} are supported by rich co-analysis and co-design tools that enable unified hardware/software implementations \cite{Vahid2002}. The answer of how an algorithm implementation can perform on hybrid \textit{HPCs} is built upon the analysis of one or all of the underlying subsystems. Indeed, the question is still on how to make adequate performance measurements in such systems.  

In computer system analysis, benchmarking is the act of measuring and evaluating the performance of computations, network processes, and connected peripherals - all under reference conditions \cite{Bouckaert2011}. A variety of benchmarks exist including Whetstone \cite{Curnow1976}, LINPAC \cite{Linpack2011}, Dhrystone \cite{weicker1984dhrystone}, Standard Performance Evaluation Corporation (SPEC) \cite{henning2000spec}, etc. Benchmarks are usually specialized; none are reported to extensively examine hybrid systems that explicitly targets hardware/software co-design.

Benchmarks can be classified into \textit{Algorithm Benchmarks} \cite{olaf2010}, \textit{Software Benchmarks} \cite{muller2003openmp}, \textit{Embedded Systems Benchmarks} and Cryptographic Benchmarks \cite{EEMBC2014}. \textit{Cryptographic Benchmarks} are available in the literature; they are designed to measure the performance of different cryptographic algorithms running under different systems \cite{menezes2010handbook}. Indeed, the use of Benchmarks is essential for performance analysis, classification, and accordingly implementation optimization. 

In this paper, we present a statistical analysis framework for performance profiling of related algorithms running under different hardware and software subsystems. The developed framework enables the deep and thorough reasoning about each hardware and software subsystem, and combines heterogeneous characteristics to provide overall rating, ranking, and classifications. The proposed framework is unique in unifying different analyses of algorithms in combined indexes. Combining analysis profiles enable the draw of conclusions on how algorithms can perform on todays hybrid processors. The proposed framework is customizable for any hybridization of processing systems and can target any model of computation or area of application. This paper includes a mathematical model for the proposed framework, a case-study from cryptography, and proposes a sample integration in development environments for hardware/software co-design. The case-study targets two high performance computing systems, namely, the \textit{Dell Precision T7500} with its \textit{dual quad-core Xeon processor} and \textit{24 GB} of \textit{RAM}, and \textit{Altera} \textit{STRATIX-II} Field Programmable Gate Array (\textit{FPGA}). The Software tools used for analysis are \textit{Quartus}, \textit{ModelSim}, and \textit{Intel VTune Amplifier}.

The paper is organized so that Section~\ref{The Analysis Framework} presents the statistical analysis framework. In Section~\ref{Case-Study}, the framework is contextualized using a case-study on cryptographic algorithms. Section~\ref{Programming Interface} presents a sample integration of the statistical framework within an integrated development environment. A thorough performance analysis and evaluation is presented in Section~\ref{Performance Analysis and Evaluation}. Section~\ref{Conclusion} concludes the paper and address possible future directions.

\section{The Analysis Framework} \label{The Analysis Framework}
The analysis framework classifies the heterogeneous sources of measurements into hardware and software analysis profiles (APs). The development of each profile includes the identification of a set of key indicators, such as speed, propagation delay, through, and power consumption. The indicators are the most extensive part of the measurement framework and should be carefully developed within the context of application. For example, for network processors, throughputs are identified as performance indicators and measured in bits-per-second; however, in graphics processors, the same indicator can be measured in frames-per-second. The measurements associated with the identified indicators may mainly quantities. The measured quantities are then each divided by similar measurements from a reference institution for normalization and for producing performance ratios. Accordingly, we can create Combined Measurement Indicators (\textit{CMIs}) using the Geometric Mean of all the calculated ratios. 

To formulate the calculation of the \textit{CMIs}, Equation ~\ref{eqn1} composes several analysis profiles:

\begin{equation} \label{eqn1}
CMI = AP_{1}\circ AP_{2}\circ...AP_{k}
\end{equation}

where $P_{k}$ is the $k^{th}$ Profile

The measurement of every Profile is done using a statistical composition of its Key Indicators (\textit{KIs}) as in Equation ~\ref{eqn2}.

\begin{equation} \label{eqn2}
P_{j} = KI_{j.1}\circ KI_{j.2}\circ KI_{j.n}
\end{equation}

where $KI_{j}$ is the $j^{th}$ Key Indicator

Therefore, The \textit{CMI} is the statistical composition of all the key indicators of all Profiles as in Equation ~\ref{eqn3}.

\begin{equation} \label{eqn3}
CMI = KI_{k.j.1}\circ KI_{k.j.2}\circ ...KI_{k.j.n}
\end{equation}

The Key Indicator values are then each divided by a reference measurements for normalization and for producing performance ratios as in Equation ~\ref{eqn4}.

\begin{equation} \label{eqn4}
ratio_{i} = \frac{KI_{k.j.i}}{KI_{k.j.i}^{ref}},
\end{equation}

where $ratio_{i}$ is the $i^{th}$ ratio, and $i \in \{1..n\}$

Then, the \textit{CMI} is the Geometric Mean of all $n$ ratios as in Equation ~\ref{eqn5}.

\begin{equation} \label{eqn5}
CMI = \sqrt{ratio_{1} \times ratio_{2}\times... ratio_{n}}
\end{equation}

The Geometric Mean is used, for the \textit{CMI}, as it is able to measure the central tendency of data values that are obtained from ratios \cite{hennessy2011computer}.

\section{A Case-Study on the Lightness of Cryptographic Ciphers} \label{Case-Study}
The presented statistical framework is contextualized by analyzing the performance of a set of lightweight cryptographic ciphers as a case-study. The aims of the case-study comprise the following:

\begin{itemize}
	\item Applying the presented framework in a computationally demanding application, such as, cryptography.
	\item Developing the key indicators for the hardware subsystem.
	\item Developing the key indicators for the software subsystem.
	\item Developing a \textit{CMI} that aids the classification of cryptographic algorithms according to their lightness; the developed \textit{CMI} is called the Lightness Indicator (\textit{LI}). 
\end{itemize}

The \textit{LI} classifies the investigated algorithms according to a combination of their software and hardware characteristics. The \textit{LI} combines several key indicators including speed, memory efficiency, hardware size, and more. The analyzed cryptographic algorithms are Skipjack \cite{biham1999cryptanalysis}, 3-WAY \cite{kelsey1997related}, XTEA \cite{andem2003cryptanalysis}, KATAN and KATANTAN \cite{de2009katan}, and Hight \cite{hong2006hight}. The reference cipher is the Advanced Encryoption Standard (AES) \cite{daemen2002design}. The literature includes a variety of implementations and performance evaluations of the addressed set of cryptographic ciphers. However, the evaluations of the targeted set of ciphers are done separately with no ground for cross-evaluation. 

The identified performance metrics of the \textit{LI} are classified into hardware and software profiles. The software profile includes the several indicators including the execution time, throughput, the total number of clock cycles per instruction, and the cash miss ratio. The 
\textbf{Execution Time} (\textit{ET}) is the time between the start and the completion of a task \cite{PattHenn2013}. The calculation of the \textit{ET} allows for the determination of the Performance according to:\\
	
	$Performance = \frac{1}{ET}$\\
	
The \textbf{Throughput} (\textit{TH}) is the total amount of work done in a given time \cite{PattHenn2013}. The \textit{TH} is application specific and could be measured, for example, in bits-per-second (bps), frames-per-second (fps), etc. The \textbf{Clock Cycle per Instruction} (\textit{CPI}) is the average number of clock cycles each instruction takes to execute. Since different instructions may take different amounts of time depending on what they do, \textit{CPI} is an average of all the instructions executed in the program \cite{PattHenn2013}. the \textbf{Cache Miss Ratio} (\textit{CMR}) is the ratio of memory accesses that cause a cache miss. The cache miss ratio of an application depends on the size of the cache. A larger cache can hold more cache lines and is therefore expected to get fewer misses \cite{PattHenn2013}.

The hardware profile comprises several indicators, namely, the execution time of the hardware implementation, throughput, propagation delay, the hardware area in number of look-up tables and logic registers, and power consumption. The \textbf{Propagation Delay} (\textit{PD}) is the time required for a signal from an input pin to propagate through combinational logic and appear at an external output pin \cite{Vahid2002}. The \textbf{Look-Up Table} (\textit{LUT}) is the number of combinational adaptive lookup tables required to implementation algorithm in hardware. The number of \textit{LUTs} is an indicator of the size of hardware in Altera devices. In other devices, the area could be measured in terms the total number of gates, logic elements, slices, etc. \textbf{Logic Registers} (\textit{LRs}) are the total number of logic registers in the design. The \textbf{Power Consumption} (\textit{PC)} is the total power consumed by developed hardware in Watts \cite{Vahid2002}.

The \textit{LI} is formulated as the composition of several assessment profiles; two for the current study. Each assessment profile is the composition of several indicators. key indicators are benchmarked against measured reference implementations to produce ratios for each measurement. Based on Equation ~\ref{eqn5}, the overall \textit{LI} is defined as the geometric mean of all the calculated ratios (See Equations ~\ref{eqn6} and ~\ref{eqn7}).

\begin{equation} \label{eqn6}
LI = \sqrt[10]{ratio_{1}\cdot ratio_{2} \cdot ratio_{3}...ratio_{l}}
\end{equation}

\noindent and hence

\begin{equation} \label{eqn7}
\textit{LI} = (\prod\limits_{i=1}^l \textit{ratio}_{i})^{\frac{1}{l}}
\end{equation}

\noindent Where $l$ is the number of ratios.

The \textit{LI} enables the classification of cryptographic algorithms according to their lightness. A higher \textit{LI} is achieved through a higher throughput, a more efficient memory performance, more compact size, less complexity, less power consumption, and less resource utilization. The master \textit{LI} formula using the developed indicators is shown in Equations ~\ref{eqn8}, ~\ref{eqnSWP}, and ~\ref{eqnHWP}. The indicators that are common to the software and hardware profiles are labeled with the profile name.

\begin{equation} \label{eqn8}
LI = \sqrt[10]{SWP \cdot HWP}
\end{equation}

\begin{equation} \label{eqnSWP} 
SWP = \frac{ET_{sw,ref}}{ET_{sw}} \cdot \frac{TH_{sw}}{TH_{sw,ref}} \cdot \frac{CPI_{ref}}{CPI} \cdot \frac{CMR_{ref}}{CMR}
\end{equation}

\begin{equation} \label{eqnHWP}
HWP = \frac{ET_{hw,ref}}{ET_{hw}} \cdot \frac{TH_{hw}}{TH_{hw,ref}} \cdot \frac{PD_{ref}}{PD} \cdot \frac{LUT_{ref}}{LUT} \cdot \frac{LR_{ref}}{LR} \cdot \frac{PC_{ref}}{PC}
\end{equation}

The derivations of Equations ~\ref{eqnSWP} and ~\ref{eqnHWP} are based on the fact that indicators are either directly or inversely proportional to the developed \textit{CMI}.

\section{Programming Interface} \label{Programming Interface}
The developed statistical framework is embedded in a sample co-design IDE. The purpose of the proposed IDE is to automate and test the connectivity to the various analysis, synthesis, and evaluation tools employed in such a hybrid framework. The IDE is implemented using \textit{Java} under \textit{Netbeans}. The used implementation and performance evaluation tools comprise \textit{Altera} \textit{Quartus} for Hardware implementation and analysis, and In\textit{tel vTune Amplifier} under \textit{Visual Studio} for Software analysis. The developed IDE connects to \textit{Altera Quartus} using the \textit{TCL} commands to synthesize and generate timing analyses, pin assignments for FPGA boards, and generate bit files to program the targeted FPGAs. The IDE connects to \textit{Intel vTune Amplifier}, using Command Line and Batch Files, to perform the software analysis and calculating the total execution time, CPI, etc. The generated Hardware and Software analysis files are exported to \textit{MS Excel} to produce the complete analysis profile and charts. 

\section{Performance Analysis and Evaluation} \label{Performance Analysis and Evaluation}
The presented statistical framework provides thorough performance analysis options for algorithms running under hybrid \textit{HPCs}. The framework structure comprises an analysis profile for every processing sub-system, key indicators for each, and a formulation that produces composite indicators. The analysis profiles serve as the performance record for one processing system; this enable deep reasoning about the performance characteristics of that processing system in particular. Looking at all the analysis profiles provide an opportunity for an evaluation based on a wider range of characteristics on more than one processing system. The composite indicators, such as the lightness indicator, provides a performance analysis summary for a desired particular property. Moreover, composite indicators aid the classification and sorting of algorithms according to a combination of heterogeneous measurements. 

\begin{figure*}
	\caption{The Lightness Indicator}\label{LI}
	\centering
	\includegraphics[width=0.75\textwidth]{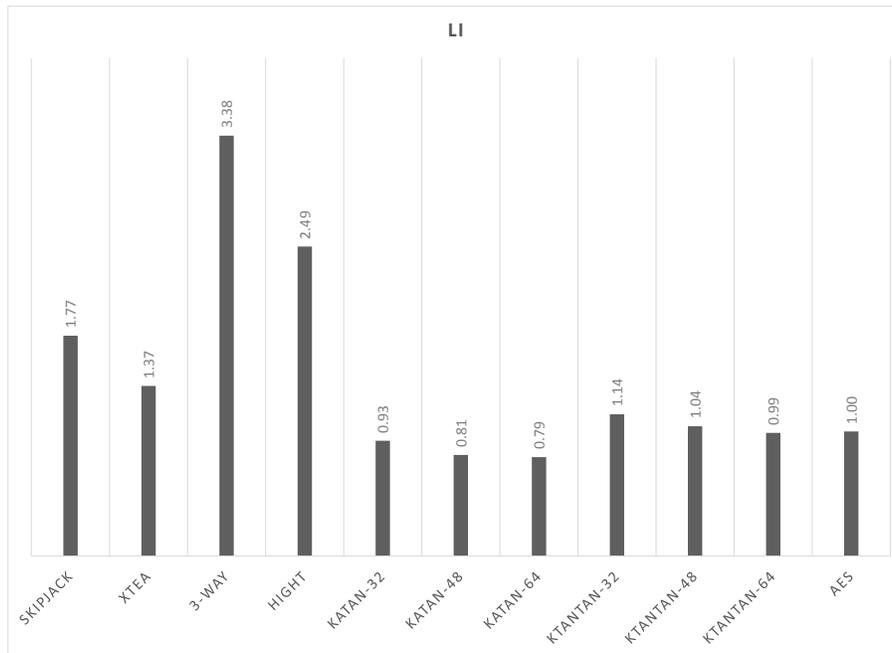}
\end{figure*}

\begin{figure}
	\caption{The Lightness Indicator; a radar chart} \label{LIRadar}
	\centering
	\includegraphics[width=0.55\textwidth]{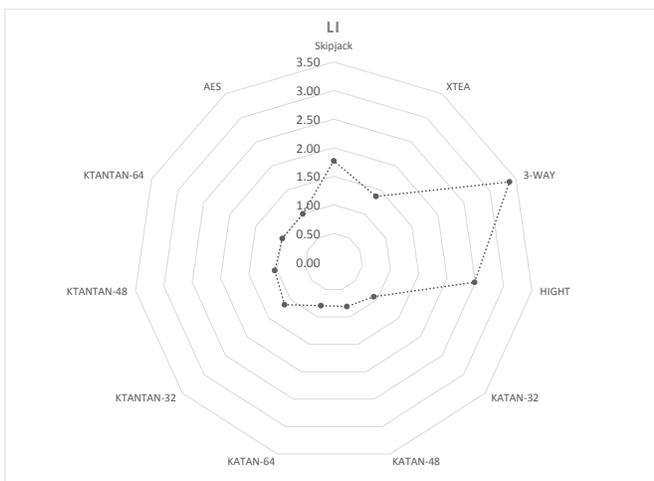}
\end{figure}

The performance of cryptographic algorithms is a primary factor in their application integration criteria. The trade-off between level of security, cost, and performance is a main issue in designing and/or analyzing lightweight ciphers. Figures \ref{LI} and \ref{LIRadar} depicts the classification of the analyzed set of algorithms according to their lightness. The algorithm that attained a larger indicator value is lighter, smaller in size, or faster than the algorithm with a lower indicator value.

The targeted set of cryptographic algorithms including Skipjack \cite{biham1999cryptanalysis}, 3-WAY \cite{kelsey1997related}, XTEA \cite{andem2003cryptanalysis}, KATAN and KATANTAN \cite{de2009katan}, and Hight \cite{hong2006hight} are all claimed to be simple, tiny, small, or lightweight. The composed \textit{LI} is built upon the presented statistical framework to provide a classification based on actual, and uniform, implementations and measurements that are based-on common grounds.

\section{Conclusion} \label{Conclusion}
In this paper, a statistical framework is developed to provide analysis options across different processing technologies. The framework classifies processing subsystems into profiles, where each can be contextualized according to a specific application. The statistical framework is adopted to investigate the lightness of a set of cryptographic algorithms that are claimed to be small in size, tiny, and efficient. The developed lightness indicator ranks the \textit{3-Way} algorithm as the lightest among all with an \textit{LI} of 3.38. \textit{Hight} achieves the second best lightness with a score of 2.49. The lowest score of 0.79 was attained by \textit{KATAN-64}. The case-study validates the statistical framework and leads to a successful classification. Future work includes the testing of reliability of the produced results through comparisons with results obtained using different methods. Future work also includes the expansion of the case-study to include additional analysis profiles and composite indicators with different performance characteristics.

\bibliographystyle{IEEETran}
\bibliography{ref}

\end{document}